\documentclass[12pt,preprintnumbers]{iopart}

\expandafter\let\csname equation*\endcsname\relax
\expandafter\let\csname endequation*\endcsname\relax

\usepackage{graphicx,amssymb,amsmath,amsthm,amsfonts,epsfig}
\usepackage[colorlinks=true]{hyperref}
\usepackage[usenames]{color}
\usepackage{epstopdf}

\usepackage{bm}
\usepackage{dcolumn}
\usepackage[utf8]{inputenc}
\usepackage{latexsym}
\usepackage{rotating}
\usepackage{xcolor}
\usepackage{longtable}
\usepackage{enumerate}
\usepackage{mathtools}
\usepackage{url}
\newcommand{\tn}{\textnormal}
\newcommand{\GSSI}{Gran Sasso Science Institute (GSSI), I-67100 L’Aquila, Italy}
\newcommand{\GranSasso}{INFN, Laboratori Nazionali del Gran Sasso, I-67100 Assergi, Italy}

\makeatletter
\renewcommand\footnoterule{%
  \kern-3\p@
  \hrule\@width2.5cm
  \kern2.6\p@}
\makeatother

\usepackage{mathrsfs}

\begin{document}

\title[]{The impact of compact binary confusion noise on tests of
  fundamental physics with next-generation gravitational-wave detectors}

\author{Luca Reali}
\address{William H. Miller III Department of Physics and Astronomy, Johns Hopkins University, Baltimore, Maryland 21218, USA}
\ead{lreali1@jhu.edu}
\author{Andrea Maselli}
\address{\GSSI}
\address{\GranSasso}
\ead{andrea.maselli@gssi.it}
\author{Emanuele Berti}
\address{William H. Miller III Department of Physics and Astronomy, Johns Hopkins University, Baltimore, Maryland 21218, USA}
\ead{berti@jhu.edu}
\vspace{10pt}
\begin{indented}
\item[]\today 
\end{indented}

\begin{abstract}

Next-generation ground-based gravitational-wave observatories such as the Einstein Telescope and Cosmic Explorer will detect $\mathcal{O}(10^{5}-10^{6})$ signals from compact binary coalescences every year, the exact number depending on uncertainties in the binary merger rate. Several overlapping signals will be present in band at any given time, generating a confusion noise background. We study how this confusion noise affects constraints on possible deviations from general relativity induced by modified gravity and environmental effects. Confusion noise impacts only the signals that last longer in band. Even for a ``golden'' GW170817-like signal, the constraints broaden by a factor in the range $[10\%, \,40\%]$ ([$70\%,\,110\%]$) for the fiducial (highest) value of the local binary neutron star merger rate. Our ability to test general relativity or constrain environmental effects will be limited by systematic errors, and not by confusion noise.

\end{abstract}

ET-0228A-23

\section{Introduction}

The LIGO~\cite{Harry:2010zz, LIGOScientific:2014pky}, Virgo~\cite{VIRGO:2014yos} and KAGRA~\cite{VIRGO:2014yos} (LVK) network of gravitational-wave (GW) interferometers has opened a new window in the study of the Universe, detecting so far $\sim 100$ signals from the coalescence of compact binaries. In the 2030s they will be joined by next-generation (XG) ground-based interferometers, Cosmic Explorer (CE)~\cite{Reitze:2019iox} in  the United States and the Einstein Telescope (ET)~\cite{Punturo:2010zz} in Europe. The increased sensitivity will provide an unprecedented redshift reach, achieving a detection rate of $\mathcal{O}(10^{5}-10^{6})$ events per year~\cite{Borhanian:2022czq,Pieroni:2022bbh,Iacovelli:2022bbs,Branchesi:2023mws}. Coupled with the routine observation of high signal-to-noise-ratio (SNR) signals~\cite{Borhanian:2022czq} , this will allow for precision tests of cosmological models~\cite{Mancarella:2022cnu,Maggiore:2019uih,Evans:2021gyd}, alternative theories of gravity~\cite{Yunes:2013dva,Berti:2015itd,Perkins:2020tra} and astrophysical scenarios of compact binary formation and evolution~\cite{Ng:2020qpk,Vitale:2018yhm}. 

The low-frequency sensitivity limit will improve to $\sim 3-7~\rm{Hz}$
down from the $10-20~\rm{Hz}$ of current detectors~\cite{Wu:2022pyg}, meaning that some GW signals will last for several hours in band~\cite{Pizzati:2021apa}. Long durations are crucial for prompt sky localization of sources to trigger electromagnetic follow-up campaigns~\cite{Ronchini:2022gwk}, but they also cause the presence of multiple overlapping signal in the data at any given time~\cite{Regimbau:2012ir,Meacher:2015rex}. 

Performing parameter estimation on coincident signals can lead to biases if the coalescence times are within less than $\sim 0.5~\rm{s}$ from each other~\cite{Pizzati:2021apa,Samajdar:2021egv,Himemoto:2021ukb,Relton:2021cax,Janquart:2022nyz}. In the same way, overlapping signals with mergers close to each other can bias parametric tests of general relativity (GR) and magnify systematic errors due to inaccurate waveform models~\cite{Hu:2022bji}.

Besides these potential biases on loud signals, the superposition of many weak, individually unresolvable signals produces a \emph{confusion noise background} component in addition to the usual instrumental noise~\cite{Crowder:2004ca,Regimbau:2009rk,Robson:2018ifk,Pozzoli:2023kxy}. In the context of XG observatories, confusion noise can reduce the redshift reach of the detectors~\cite{Wu:2022pyg} and broaden the errors on the inferred parameters of the longest signals~\cite{Reali:2022aps}.

In this work, we assess the impact of confusion noise on tests of fundamental physics in XG detectors, considering both parametric tests of GR~\cite{Yunes:2009ke} and constraints on environmental effects~\cite{Barausse:2014tra}. We generate the confusion noise from a catalog of unresolved binary neutron star (BNS) signals from state-of-the-art population models~\cite{Farrow:2019xnc,KAGRA:2021duu}. We employ the formalism of Ref.~\cite{Reali:2022aps} to compute the confusion noise power spectral density (PSD). We model environmental effects and beyond-GR corrections by adding parametrized deviations to the gravitational phase at various post-Newtonian (PN) orders~\cite{Yunes:2009ke,Cardoso:2019rou}. We then estimate the impact of confusion noise on the constraints on such deviations with an information-matrix formalism~\cite{Finn:1992wt}.

The paper is organized as follows. In Sec.~\ref{sec:confnoise} we describe how we generate and characterize the confusion noise. In Sec.~\ref{sec:numsetup} we summarize the parametrized post-Einsteinian (ppE) framework we use to constrain environmental effects and deviations from GR. In Sec.~\ref{sec:results} we present our results.

Throughout this work, we adopt the $\Lambda$CDM cosmological model with parameters taken from Planck 2018~\cite{Planck:2018vyg}.

\section{Confusion noise}
\label{sec:confnoise}

\subsection{Theoretical framework}

Let us consider a loud, detected GW signal $h(\vec{\xi})$ with true parameters $\vec{\xi}$. If $N$ unresolved signals $\{h_{\rm over}^j\}_{j=1}^N$ are present in band at the same time, the detector output reads 
\begin{equation}
{s}(t) = {h}(t;\vec{\xi}) + {n}(t) + \Delta {H}(t) \,,
\label{math:detoutput}
\end{equation}
where $n$ is the instrumental noise, and $\Delta {H}$ is the confusion noise produced by the superposition of the $N$ overlapping signals~\cite{Antonelli:2021vwg}:
\begin{equation}
\Delta H(t) = \sum_{j=1}^{N} h^j_{\rm over}(t)\,.
\label{math:confnoise}
\end{equation}
The number of coincident background signals (and thus the confusion noise) is mainly determined by two quantities: the duration in band of the resolved signal $h$, and the merger rate $R_0$ of the astrophysical population generating the unresolved signals~\cite{Reali:2022aps}. At leading (quadrupolar) order, the duration in band of a detected, inspiral-dominated GW signal is given by
\begin{equation}
T_{\rm det} = \frac{5}{256}\left(\frac{G\mathcal{M}_{\rm z}}{c^3}\right)^{-5/3}(\pi f_0)^{-8/3} \,.
\label{eq:duration}
\end{equation}
Here, $f_0$ is the low-frequency sensitivity limit of the detector and $\mathcal{M}_{\rm z}$ is the detector-frame chirp mass of the source
\begin{equation}
\mathcal{M}_{\rm z} = (1+z)\,\frac{(m_1 m_2)^{3/5}}{(m_1+m_2)^{1/5}}\,,
\label{math:chirpmass}
\end{equation}
where $z$ is the redshift and $m_{1,2}$ denotes the binary component masses. For a given detected signal $h$, smaller values of the redshifted chirp mass $\mathcal{M}_{\rm z}$ correspond to a larger number of coincident background signals $N$, and to a larger impact of the confusion noise on parameter estimation errors~\cite{Reali:2022aps}.

Assuming that the confusion noise is Gaussian and stationary, it can be modeled by a power spectral density (PSD) $S_{\rm conf}$ via~\cite{Antonelli:2021vwg,Reali:2022aps}
\begin{equation}
\langle \Delta\tilde{H}(f)\Delta\tilde{H}^*(f') \rangle = \frac{1}{2}\delta(f-f')S_{\rm conf}(f) \,,
\label{math:psdconf}
\end{equation}
where we denote the Fourier transform with a tilde, complex conjugates with a star, and ensemble averages with $\langle \cdot \rangle$. For a background of inspiral-dominated GW signals from compact binaries, the confusion noise PSD can be approximated by a power law~\cite{Reali:2022aps}
\begin{equation}
S_{\rm conf}(f)=A_{\rm ref}\,(f/f_{\rm ref})^{-7/3}\,,
\label{math:powerlawpsd}
\end{equation}
with $f_{\rm ref}$ an arbitrary reference frequency. The total PSD for each detector in the presence of confusion noise is then given by
\begin{equation}
S_{\rm tot}(f) = S_{\rm n}(f) + S_{\rm conf}(f)\,,
\label{math:totpsd}
\end{equation}
where $S_{\rm n}$ is the instrumental-noise PSD.

\subsection{Generation of confusion noise}

To generate the confusion noise, we consider a background of BNS signals, which is expected to be the dominant GW background from compact binaries in XG detectors~\cite{Zhou:2022nmt,Sachdev:2020bkk}. Furthermore, BNS signals last longer in band compared to binary black hole (BBH) or neutron star-black hole (NSBH) signals, meaning that they have a higher chance of overlap~\cite{Pizzati:2021apa}.

We adopt a BNS population consistent with the latest LVK GWTC-3 catalog~\cite{KAGRA:2021duu}. The component masses are sampled according to the preferred model of Ref.~\cite{Farrow:2019xnc}: the primary mass $m_1$ follows a double Gaussian distribution with means $1.34\,M_{\odot}$ and $1.47\,M_{\odot}$, standard deviations $0.02\,M_{\odot}$ and $0.15\,M_{\odot}$, and mixing fraction $0.68$; the secondary mass is instead distributed uniformly within the range $m_2\in[1.14,1.46]\,M_{\odot}$. We assume nonspinning BNSs and neglect tidal deformabilities. For the BNS redshift $z$, we adopt the same distribution as Ref.~\cite{Borhanian:2022czq}. We assume that the binary formation rate follows the Madau-Dickinson~\cite{Madau:2014bja} cosmic star formation rate, and we obtain the merger rate by convolving the SFR with a standard $p(t_d)\propto 1/t_d$ time-delay distribution~\cite{Dominik:2013tma,Meacher:2015iua,LIGOScientific:2017zlf}. 

The normalization of the merger rate $R_0$ is set by the measured local merger rate from LVK observations~\cite{KAGRA:2021duu}. We choose a fiducial value of $R_0 = 320\,\rm Gpc^{-3}yr^{-1}$, which is consistent with the estimates of both the GWTC-2 and GWTC-3 catalogs~\cite{LIGOScientific:2020kqk,KAGRA:2021duu}. To characterize the impact of the large astrophysical uncertainty on the local merger rate, we vary $R_0$ within the $90\%$ confidence interval from the GWTC-3 catalog~\cite{KAGRA:2021duu}, going from a minimum value $R_0^{\rm low}=10\,\rm Gpc^{-3}yr^{-1}$ to a maximum of $R_0^{\rm high}=1700\,\rm Gpc^{-3}yr^{-1}$.

For each BNS sampled from our population model, we generate a GW signal and assess detectability by computing its network SNR. We consider a detector network of three XG ground-based observatories. We choose one CE detector with $40$-km arm length in the US~\cite{Reitze:2019iox}, one CE with $20$-km arm length in Australia~\cite{Reitze:2019iox}, and one ET in Italy~\cite{Punturo:2010zz}. We set a detectability threshold of $\rho_{\rm thr}=12$ and assume that only signals with SNR lower than $\rho_{\rm tr}$ contribute to the background. We assume $f_0=3~\rm{Hz}$ to be the low-frequency sensitivity limit of every detector in our network.

We assign a fixed time of arrival $t_0$ to the resolved signal $h$, and uniformly sample the arrival times of the background signals around $t_0$. Then, we associate  an interval in time domain to every signal in our catalog from its time of arrival and duration in band. If the interval associated to a background signal overlaps with the interval associated to $h$, the background signal contributes to the confusion noise (see Ref.~\cite{Reali:2022aps} for more details). 

The duration of each signal in band is computed at $3.5$ post-Newtonian (PN) order with the public package \textsc{PyCBC}~\cite{Usman:2015kfa}. GW signals are generated with the inspiral-only waveform model \texttt{TaylorF2}~\cite{Sathyaprakash:1991mt,Buonanno:2009zt} and SNRs are computed with the public package \textsc{gwbench}~\cite{Borhanian:2020ypi}.

\section{Numerical setup}
\label{sec:numsetup}

We investigate the impact of confusion noise on fundamental physics using a Fisher information matrix formalism~\cite{Vallisneri:2007ev}. We assume that the generic GW signal in the frequency domain $\tilde{h}(f,\vec{\theta})$, which depends on the source parameters $\vec{\theta}$, is observed by XG detectors with large SNRs. 
In this regime, given the interferometer's output $s$, we expect the parameters $\vec{\theta}$ to be strongly peaked around their true values $\vec{\xi}$, so their likelihood function can be described by a gaussian distribution and the posterior reads
\begin{equation}\label{prob1}
p( \vec{\theta}\vert s)\propto p^{(0)}(\vec{\theta})e^{-\frac{1}{2}(\theta^{i}-\xi^{i})\Gamma_{ij}(\theta^{i}-\xi^{i})^\tn{T}}\ ,
\end{equation}
where $p^{(0)}(\vec{\theta})$ is the prior on $\vec{\theta}$, and $\Gamma_{ij}$ are the elements of the Fisher matrix:
\begin{equation}
\Gamma_{ij}=\left(\frac{\partial h}{\partial \theta^i}\bigg\vert \frac{\partial h}{\partial \theta^j}\right)\bigg\vert_{\vec{\theta}=\vec{\xi}}\ . \label{math:FisherM}
\end{equation}
Here we have defined the inner product
\begin{equation}
(h_1\vert h_2)=2\int_{f_\tn{min}}^{f_\tn{max}}\frac{\tilde{h}_1(f) \tilde{h}_2^{\star}(f)+
\tilde{h}_1^{\star}(f) \tilde{h}_2(f)}{S(f)}df\ ,\label{math:inner}
\end{equation} 
where $S(f)$ is the noise spectral density of the detector \cite{Cutler:1994ys}.
The covariance matrix of the parameters, $\Sigma^{ij}$, is simply given by $\Sigma^{ij}=\left(\Gamma^{-1}\right)^{ij}$. 
Using Eq.~\eqref{math:inner} we can define the SNR of the signal $\tilde{h}(f)$ as
\begin{equation}
\rho=(h\vert h)^{1/2}\ .\label{math:snr}
\end{equation}
In the following we fix $f_\tn{min}=3$~Hz for both ET and CE, while $f_\tn{max}$ is conventionally set the to innermost stable circular orbit (ISCO) frequency of the Schwarzschild spacetime, $\pi Mf_\tn{max}=6^{-3/2}$, with $M$ total mass of the system (see e.g.~\cite{Cutler:1994ys,Poisson:1995ef,Berti:2004bd}).

We use as a baseline waveform model the Fourier-domain TaylorF2 approximant~\cite{Sathyaprakash:1991mt,Damour:2000gg}, which describes the inspiral phase of the binary:
\begin{equation}
\tilde h (f) = \Omega{\cal A}_\tn{N}(f) \textnormal e^{i \phi  (f)}\ .\label{math:waveform}
\end{equation}
The phase $\phi(f)$ is given by a post-Newtonian (PN) expansion:
\begin{align}
\phi(f)=2\pi ft_c-\phi_c-\frac{\pi}{4}+\frac{5}{128({\pi \cal M}f)^{5/3}}\bigg[&\sum_{i=0}^7\psi_\tn{pp}^{(i/2)}(\pi M f)^{i/3}\nonumber\\
&+\phi_\tn{T}^{(5)}(\pi M f)^{10/3}
+\phi_\tn{T}^{(6)}(\pi M f)^{12/3}\bigg]\ ,\label{math:GWphase}
\end{align}
where a term proportional to $(\pi {\cal M} f)^{i/3}$ is said to be of
$i$th PN order, and $(t_c,\phi_c)$ are the time and phase at coalescence. The phase \eqref{math:GWphase} contains: (i) point-particle (pp) terms up to the 3.5PN $(i=7)$ order~\cite{Damour:2000gg, Arun:2004hn, Buonanno:2009zt, Abdelsalhin:2018reg}, which depend on the binary chirp mass $\mathcal{M} = (m_1 m_2)^{3/5}/(m_1+m_2)^{1/5}$ and on the symmetric mass ratio $\eta = m_1 m_2/(m_1+m_2)^2$; (ii) linear spin corrections proportional to the (anti)symmetric combinations of the spin parameters $\chi_{s}=(\chi_{1}+\chi_{2})/2$ and $\chi_{a}=(\chi_{1}-\chi_{2})/2$ up to 3PN $(i=6)$ order, as well as quadratic-in-spin corrections entering at 2PN $(i=4)$ order\footnote{We neglect here the contribution induced by spin-induced quadrupole moments, which appears at 2PN order in $\psi(f)$.}~\cite{Krishnendu:2017shb}; (iii) 5PN and 6PN tidal 
terms, $\psi_\tn{T}^{(5)}$ and  $\psi_\tn{T}^{(5)}$, which depend on the effective tidal parameter $\tilde \Lambda$\footnote{The 6PN coefficient contains a second tidal parameter, $\delta\tilde{\Lambda}$, which is in general very small for realistic equations of state ($\delta\tilde{\Lambda}\sim0$), and therefore will be neglected.}~\cite{Flanagan:2007ix, Vines:2011ud,Lackey:2014fwa}. 
We truncate the waveform amplitude at the leading (Newtonian) order:
\begin{equation}
\mathcal{A}_\textnormal{N} = \sqrt{\frac{5}{24}} 
\frac{\mathcal{M}^{5/6}f^{-7/6}}{\pi^{2/3}d_L}\ ,
\end{equation}
where $d_L$ is the luminosity distance. The geometric factor $\Omega$ in Eq.~\eqref{math:waveform} depends on the inclination angle $\iota$ between the line of sight of the source and its orbital angular momentum and on the detectors' antenna pattern functions, i.e., it is a function of the source position in the sky $(\theta, \varphi)$ and of the polarization angle $\psi$. Here we consider two different scenarios: in the first we average over all binary orientations, such that the waveform is specified by the parameters $\vec{\theta}=({\cal A}_\tn{N},{\cal M},\eta,\chi_s,\chi_a,\tilde{\Lambda},t_c,\phi_c)$, which yield a $8\times 8$ covariance matrix; in the second, we include in the Fisher matrix \eqref{math:FisherM} all of the angles that determine the geometric factor $\Omega$ , i.e., $\vec{\theta}=({\cal A}_\tn{N},{\cal M},\eta,\chi_s,\chi_a,\tilde{\Lambda},t_c,\phi_c,\iota,\theta,\varphi,\psi)$.

We augment the waveform \eqref{math:waveform} by including either:

\begin{itemize}
\item[(i)] a beyond-GR parametric deviation in the PN expansion of the phase $\phi(f)$, or
\item[(ii)] a phase shift which encodes the presence of environmental effects.
\end{itemize}

In the first case, following Ref.~\cite{Yunes:2009ke}, we modify the GR term by an additive term:
\begin{equation}
\phi(f)\ \longrightarrow \ \phi(f)+\beta({\cal M} \pi f)^{(2\gamma-5)/3}\ ,\label{math:ppE}
\end{equation}
where $\gamma$ identifies the leading PN order of the non-GR contribution, and $\beta$ is a coefficient that depends on the theory and (possibly) on the binary parameters.

Environmental effects can be included in the TaylorF2 waveform with a similar methodology. We focus here on three different phenomenological terms that correspond to gravitational pull, gravitational drag, and collisionless accretion~\cite{Cardoso:2019rou}, which modify Eq.~\eqref{math:GWphase} as follows:
\begin{equation}
\phi(f)\ \longrightarrow \ \phi(f)+
\rho_0\kappa M^2(M f)^{-\delta}
\quad  
\begin{cases}
\delta_\tn{pull}=2\quad &\kappa_\tn{pull}= 1 \\
\delta_\tn{drag}=11/3\quad &\kappa_\tn{drag}= -\eta^{-3}(1-3\eta)\pi^{-11/3} \\
\delta_\tn{accretion}=3\quad &\kappa_\tn{accretion}= -\eta^{-1}\pi^{-3}
\end{cases}\ ,
\label{math:enveffects}
\end{equation}
where $\rho_0$ is the average density of the medium in which the binary system evolves.

In both cases (i) and (ii), the parameter vector $\vec{\theta}$ includes an extra term, corresponding to the parameter $\beta$ in case (i) and to the parameter $\rho_0$ in case (ii).

We perform our Fisher analysis on four representative systems with parameters similar to the two BNS systems GW170817~\cite{LIGOScientific:2017vwq}, GW190425~\cite{LIGOScientific:2020aai}, to the NSBH binary GW200115~\cite{LIGOScientific:2021qlt}, and to the BBH GW150914~\cite{LIGOScientific:2016aoc}, respectively. The masses, luminosity distance and tidal parameters of these sources are listed in Table~\ref{tab:parinj}.  For all systems we assume negligible spins ($\chi_1=\chi_2=0$), and we also set $t_c=\psi_c=0$. Finally, for the BBH event GW150914 we remove the tidal deformability from the list of waveform parameters, because tidal deformabilities are equal to zero for 
black holes \cite{Damour:2009vw,Binnington:2009bb,Pani:2015hfa, Gurlebeck:2015xpa, LeTiec:2020bos,LeTiec:2020spy}.

\begin{table}[htbp!]
\centering
\begin{tabular}{ccccc}
\hline
system &$m_1 [M_\odot]$ & $m_2 [M_\odot]$ &$d_L [\tn{Mpc}]$ &  $\tilde{\Lambda}$   \\ 
\hline
GW170817 & 1.46 & 1.27 & 40.1 & 600\\
GW190425 &2 & 1.4 & 159 & 500\\
GW200115 & 5.7 & 1.5 & 326 & 56\\
GW150914 & 36 & 30 & 475 & -\\
\hline
\hline
\end{tabular}
\caption{Injected values of the component masses, luminosity distance and effective tidal deformability 
for the four systems considered in this work.}\label{tab:parinj}
\end{table}

\begin{figure}[htbp!]
\centering		
\includegraphics[width=1\textwidth]{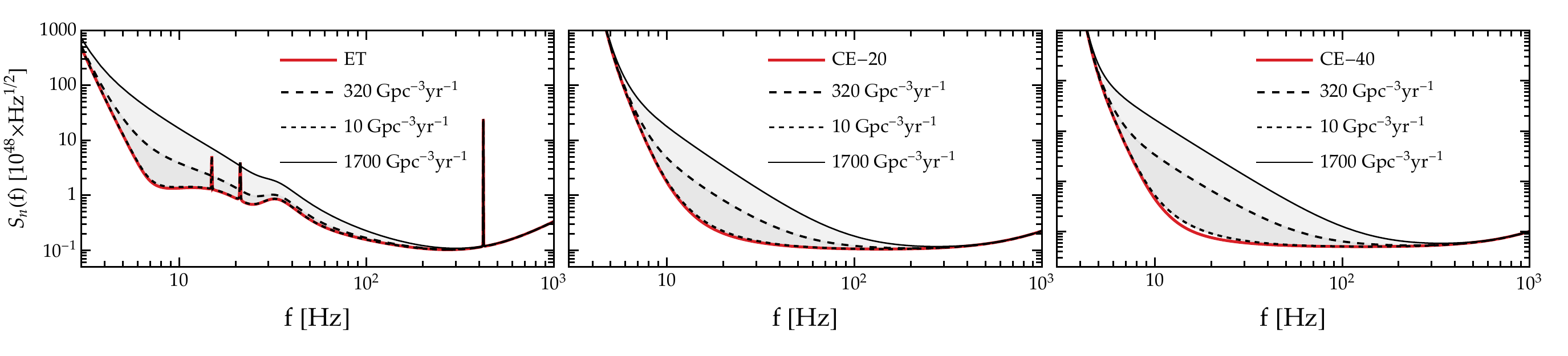}
\includegraphics[width=1\textwidth]{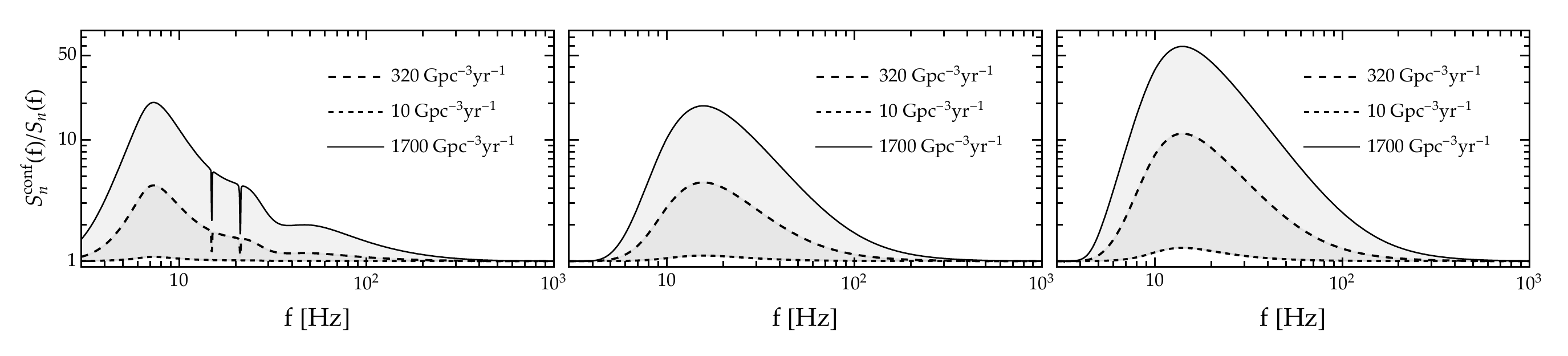}
\caption{Top row: Noise PSD for the three GW detectors considered in this work (from left to right: ET, CE-20 and CE-40).  For each interferometer we show the noise PSD at design sensitivity (solid red curve), as well as the PSDs obtained by adding the contribution of the confusion noise, assuming a binary system with source parameters similar to GW170817. The shaded region represents the range compatible with current uncertainties in the local merger rate, with the upper (solid black line) and lower (dotted black line) edges corresponding to ${\cal R}_0=1700\tn{ Gpc}^{-3}\tn{yr}^{-1}$ and ${\cal R}_0=10\tn{ Gpc}^{-3}\tn{yr}^{-1}$, respectively.  Bottom row: Ratio between confusion noise and instrumental noise PSDs.}
    \label{fig:noise}
\end{figure} 

\section{Results}
\label{sec:results}

\begin{figure}[htbp!]
\centering		
\includegraphics[width=0.6\textwidth]{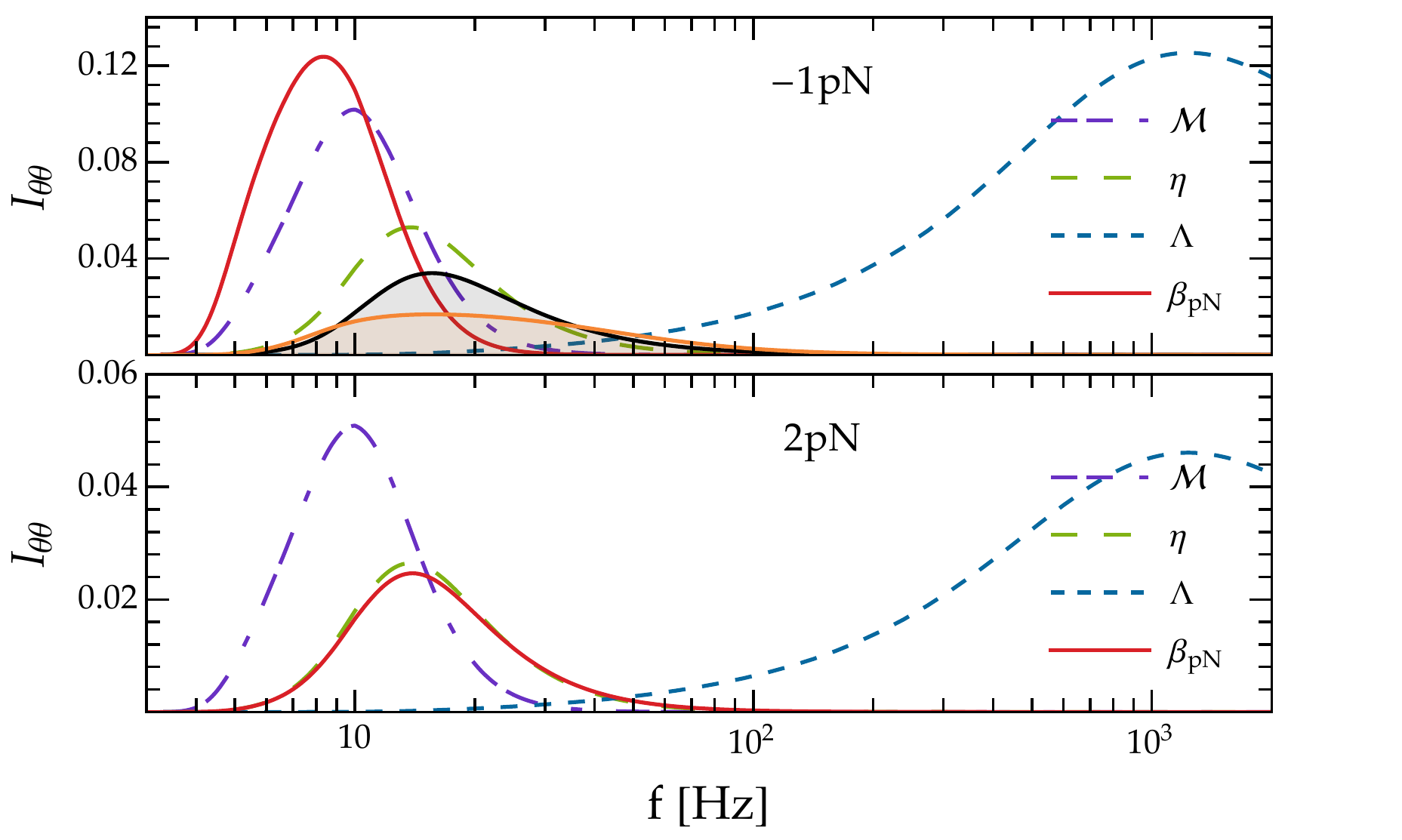}
\caption{Numerical value of the integrand in Eq.~\eqref{math:inner} as a function of the frequency for various parameters of interest. The top and bottom panels refer to two different non-GR modifications: the top panel is a -1PN (dipolar) term, of the kind that would appear e.g. in some scalar-tensor theories of gravity, while the bottom panel refers to a 2PN order modification. For illustration, here we focus on a GW170817-like binary observed by CE20.}
\label{fig:measurability}
\end{figure}

Figure~\ref{fig:noise} shows the total PSD $S_{\rm tot}(f)$ (top panels) and the ratio between confusion and instrumental noise PSDs $S_{\rm n}^{\rm conf}(f)/S_{\rm n}(f)$ for the longest signal we consider, a GW170817-like BNS. Confusion noise dominates over instrumental noise at low frequencies, with the ratio between the PSDs peaking between $7$ and $12$ Hz for all detectors. Given the present uncertainty in the BNS merger rate, the confusion noise PSD ranges from being comparable to the instrumental noise PSD at the lowest end of the estimated merger rate, to being $\sim 20$ or even $\sim 50$ times larger than the instrumental noise (in the worst-case scenario) for the highest estimated merger rates.

We can understand which parameter errors are most affected by confusion noise by comparing the PSD ratio $S_{\rm conf}(f)/S_{\rm n}(f)$ with the (normalized) measurability integrand $I_{\theta\theta}$~\cite{Damour:2012yf}. This quantity is defined as the integrand of the corresponding diagonal element of the Fisher matrix, and it illustrates what range of frequencies is most important for the measurability of the given parameter. In Fig.~\ref{fig:measurability} we plot the measurability integrands for selected binary parameters and for two different beyond-GR modifications of the waveform, entering either at -1PN (top panel) or 2PN (bottom panel). In both cases, constraints on the deviation parameters $\beta_{\rm PN}$ rely on measurements at frequencies close to the peak of PSD ratio $S_{\rm conf}/S_{\rm n}$. Therefore we can expect confusion noise to broaden the constraints on these parameters, at least for long enough signals.

\begin{figure}[t]
  \centering \includegraphics[width=0.45\textwidth]{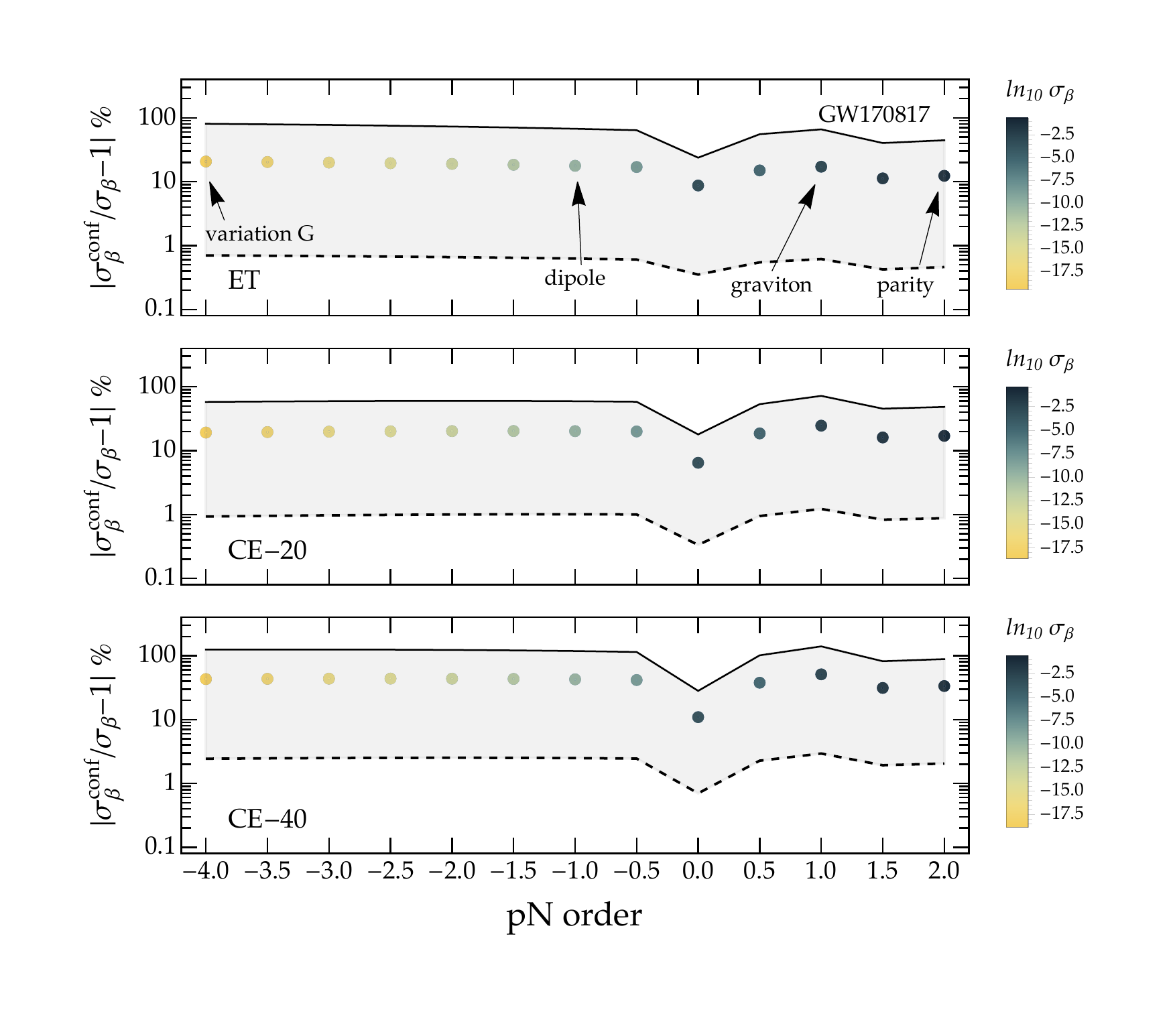} \includegraphics[width=0.45\textwidth]{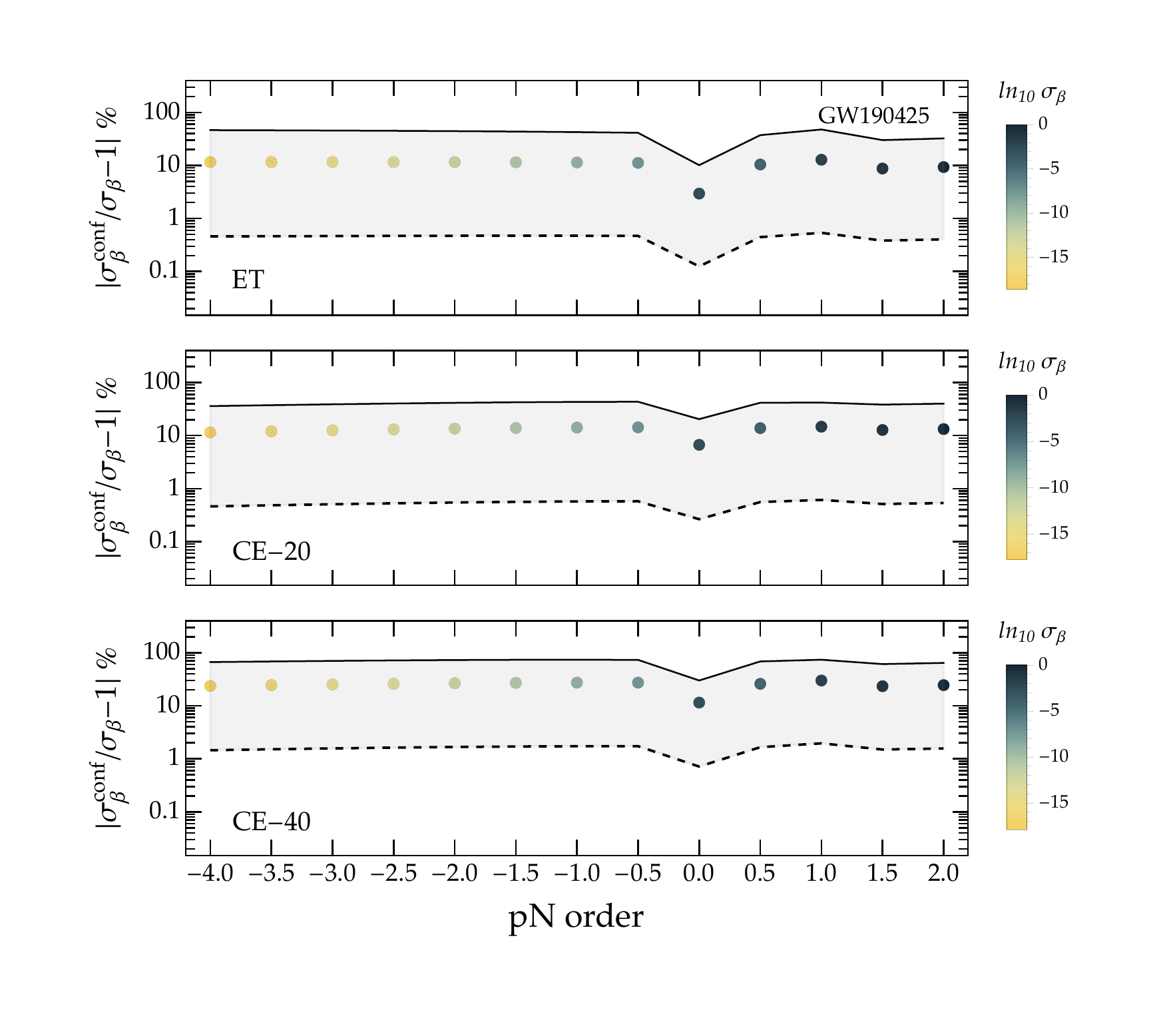}
  \caption{Relative percentage error on the ppE parameter $\beta$ which controls the amplitude of the beyond-GR correction to the waveform phase [see Eq.~\eqref{math:ppE}] for three different detectors (top to bottom: ET, CE20 and CE40), and for a binary merger similar to GW170817 (left) or GW190425 (right). All of these results are obtained by averaging over the inclination angle, polarization angle, and sky location. Arrows in the top-left panel illustrate the mapping between specific PN orders and certain interesting physical modifications to the waveform. The color code of the filled circles at fixed PN orders correspond to the error on the parameter $\beta$ that would be achievable in the absence of confusion noise. The shaded regions at each PN order measure the percentage broadening of this constraint due to confusion noise: the upper (lower) values of these broadenings correspond to high (low) coalescence rate, while the bullet points were computed for the current best estimate of the local BNS rate (${\cal R}_0=320\tn{ Gpc}^{-3}\tn{yr}^{-1}$).}
\label{fig:ppE}
\end{figure}

Figure~\ref{fig:ppE} quantifies the extent of this broadening for the two longest detected signals in our study: the GW170817-like and GW190425-like BNS systems. As a general trend, confusion noise-induced broadening is slightly less important for higher-order PN corrections, which are mostly measured at higher frequencies. However, this trend is barely noticeable. The confusion noise-induced broadening for the fiducial value of the local BNS merger rate (${\cal R}_0=320\tn{ Gpc}^{-3}\tn{yr}^{-1}$) is shown by bullets: for all three detectors, constraints on the deviation parameters broaden by a factor ranging between $\sim 10\%$ and $\sim 50\%$. The worst-case scenario corresponds, of course, to the highest merger rate (solid black lines marking the top edge of the gray band in each panel). In this case, the broadening ranges between $50$ and $110\%$. As expected, the broadening is negligible at the lowest end of the estimated merger rate.

We do not show plots for the GW200115-like NSBH binary, because they have the same qualitative behavior shown in Fig.~\ref{fig:ppE}. The impact of confusion noise is negligible for shorter GW signals: our Fisher analysis for a GW150914-like system shows that the constraints on the ppE parameters are almost unaffected, which changes smaller than $1\%$ at all pN orders.

\begin{figure}[htbp!]
  \centering		
  \includegraphics[width=0.4\textwidth]{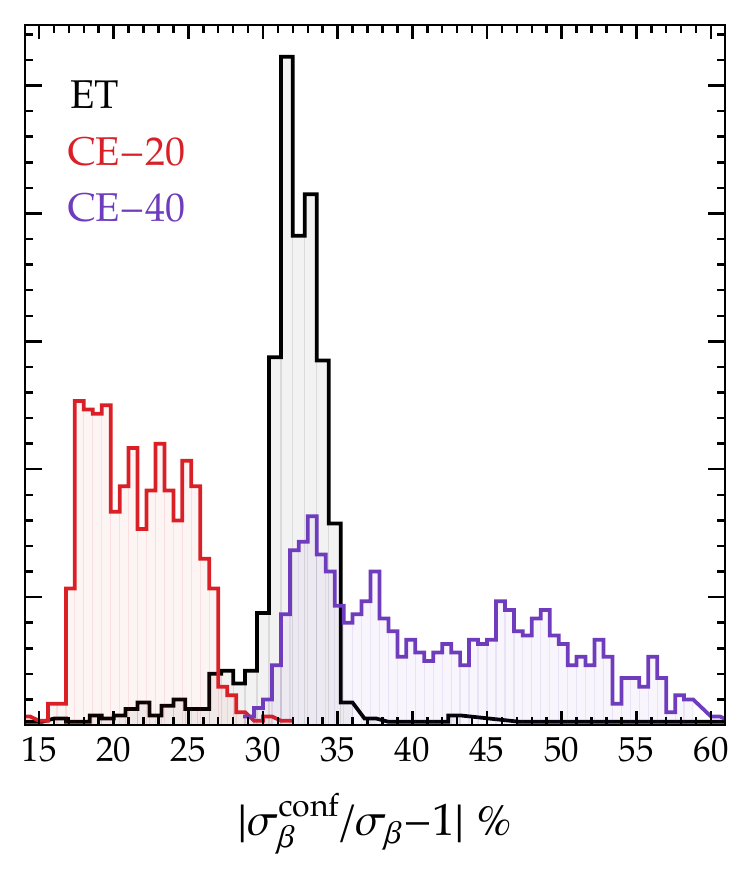}
  \caption{Distribution of the broadening in the ppE amplitude $\beta$ for a -1PN correction to the waveform phase. Each histogram is obtained by drawing $10^3$ angles from uniform distributions, with $\cos\iota\sim{\cal U}_{[-1,1]}$, $\cos\theta\sim{\cal U}_{[-1,1]}$, $\varphi\sim{\cal U}_{[0,2\pi]}$ and $\psi\sim{\cal U}_{[0,\pi]}$.}
\label{fig:randomangles}
\end{figure}

The constraints shown in Fig.~\ref{fig:ppE} are computed by averaging over inclination angle, polarization angle and sky location. In Fig.~\ref{fig:randomangles} we show how the broadening in $\beta_{\rm PN}$ for different detectors changes as we sample isotropically over these angular parameters. For illustration, we focus on a GW170817-like binary, the fiducial local BNS merger rate, and a dipolar (-1PN) deviation from GR. The qualitative shape of the histograms depends on the detector's location, orientation and antenna pattern. The variability is largest for CE40: in this case, the percentage broadening on $\beta_{\rm PN}$ ranges between $\sim 30\%$ and $\sim 60\%$.

\begin{figure}[htbp!]
  \centering
  \includegraphics[width=0.5\textwidth]{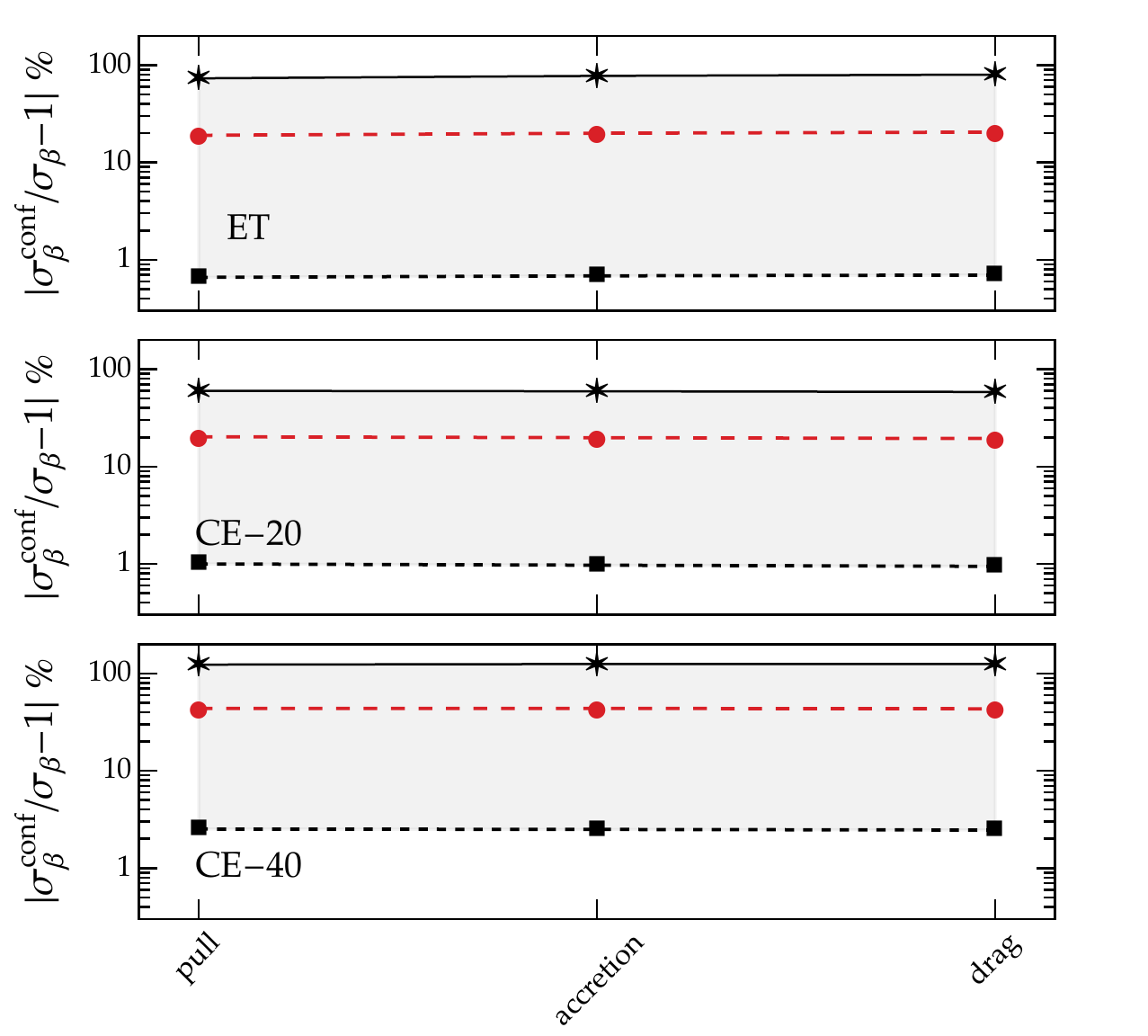}
  \caption{Relative percentage error on the matter density $\rho_0$ inferred by introducing different environmental effects within the GW phase (namely, gravitational pull, accretion, and gravitational drag). The results shown here refer to a GW170817-like BNS source.}
\label{fig:env}
\end{figure}

Finally, in Fig.~\ref{fig:env} we present results similar to Fig.~\ref{fig:ppE}, but focusing on environmental effects. More specifically, we show how confusion noise would affect constraints on the matter density $\rho_0$ for the three different phenomenological modifications introduced in Eq.~\eqref{math:enveffects} in the ``worst-case'' scenario of a GW170817-like resolved signal. The broadening is mildly detector-dependent: it ranges from $\sim 20\%$ to $\sim40\%$ at the fiducial local BNS merger rate, from $\sim 70\%$ to $\sim 110\%$ at the highest end of the merger rate, and (as usual) it becomes negligible at the lowest end of the local merger rate.

\section{Conclusions}
\label{sec:conclusions}

We have studied how confusion noise affects the constraints that XG ground-based detectors, such as ET and CE, may place on beyond-GR parametric deviations and environmental effects.
We have found that the contribution of confusion noise is negligible for most signals, becoming relevant only for the longest signals in band, such as low-redshift BNS signals. Even in the worst-case scenario (a ``golden'' GW170817-like BNS at $\sim 40~\rm{Mpc}$), the constraints on beyond-GR deviations (or on the average density of the medium interacting with the binary) broaden by a modest amount: $\sim 10-40\%$ for our fiducial BNS merger rate, and $\sim 70-110\%$ at the highest end of the local merger rates compatible with current LVK observations. It may be possible to further reduce the impact of confusion noise on parameter estimation by more sophisticated data analysis techniques, such as global fitting methods~\cite{Littenberg:2020bxy,Robson:2017ayy,Petiteau:2012zq} or simultaneous fitting of the foreground and background parameters~\cite{Biscoveanu:2020gds}. The exploration of these methods is an interesting topic for future work.

The estimates provided in this work are somewhat conservative, because we only considered unresolved signals from BNSs (which, however, are expected to dominate the GW background from compact binaries). The inclusion of unresolved signals from NSBHs and BBHs would increase the confusion noise by a factor of order unity and introduce non-Gaussian components that complicate the analysis, but it would not change our main conclusions.

An additional caveat to consider is the low-frequency sensitivity limit of the detectors. Throughout our study, we set the limit to $3$~Hz for all of the interferometers. Decreasing this limit would increase the duration in band of all the GW signals, and thus the number of overlapping signals at any given time.

An important implication of our work is that the limiting factor in our ability to test GR or constrain environmental effects will be systematic errors due to our imperfect modeling of GR waveforms, and not confusion noise.

\section*{Acknowledgments}

E.B. and L.R. are supported by NSF Grants No. AST-2006538, PHY-2207502, PHY-090003 and PHY-20043, and NASA Grants No. 20-LPS20-0011 and 21-ATP21-0010. 
A.M. and E.B. acknowledge support from the ITA-USA Science and Technology Cooperation programme (CUP: D13C23000290001), supported by the Ministry of Foreign Affairs of Italy (MAECI).
A.M. acknowledges financial support from the Italian Ministry of University and Research (MUR) for the PRIN grant METE under contract no. 2020KB33TP.
This research project was conducted using computational resources at the Maryland Advanced Research Computing Center (MARCC). 
The authors acknowledge the Texas Advanced Computing Center (TACC) at The University of Texas at Austin for providing {HPC, visualization, database, or grid} resources that have contributed to the results reported in this paper \cite{10.1145/3311790.3396656}. URL: http://www.tacc.utexas.edu. 

\clearpage

\bibliographystyle{iopart-num.bst}
\bibliography{main}

\end{document}